# CT-CFAR: A Robust CFAR Detector Based on CLEAN and Truncated Statistics in Sidelobe-Contaminated Environments

Jiachen Zhu, Fangjiong Chen, *Member*, *IEEE*, Jie Wu, and Ming Xia

*Abstract*—This paper proposes a constant false alarm rate (CFAR) target detection algorithm based on the CLEAN concept and truncated statistics to mitigate the non-homogeneity of reference samples caused by sidelobe contamination and other abnormal interferences within the reference window. The proposed algorithm employs truncated statistics to separate target and noise components in the radar echo power spectrum, thereby restoring the homogeneity assumption of the reference window. In addition, learnable historical sidelobe information is introduced to enhance the robustness and environmental adaptability of the detection process. Furthermore, based on multichannel echo data, a target reconstruction model that combines the Candan algorithm with least-squares estimation is established, incorporating the CLEAN concept to suppress sidelobe interference. Monte Carlo simulations and real-world measurement experiments demonstrate that the proposed CT-CFAR algorithm achieves high-precision target detection without requiring prior knowledge of abnormal samples. Compared with various CFAR algorithms, the proposed approach overcomes the limitations of the reference window, accurately estimates the noise spectrum, and exhibits superior detection performance and computational efficiency in complex scenarios affected by sidelobe contamination.

## I. INTRODUCTION

With the continuous advancement of radar technology [1], radar systems are gradually evolving from their traditional attributes of being bulky and costly toward miniaturization, affordability, and widespread accessibility [2]. This evolution has significantly promoted the pervasive deployment of radar technology in everyday life [3].

Among various radar architectures, frequency-modulated continuous-wave (FMCW) radar has garnered significant attention for its high range resolution, robust penetration capability, and notable resistance to environmental interference [4-6]. Consequently, it has been extensively deployed in a diverse range of applications, such as vital sign monitoring [7, 8], human activity recognition [9, 10], autonomous driving [11, 12], security surveillance [13, 14], and post-disaster search and rescue [15]. As radar applications continue to expand into increasingly complex and dynamic environments, achieving reliable and high-precision target detection in multi-target and sidelobe-contaminated environments has become a critical challenge that directly impacts overall system performance [16].

Currently, the most widely adopted target detection algorithm for FMCW radar is the constant false alarm rate (CFAR) detector based on a sliding reference window [17]. This algorithm adaptively estimates the detection threshold from the observations within the reference window to distinguish targets from background noise. The primary distinction among various CFAR schemes lies in their methodology for estimating the local noise power. In homogeneous environments, the cell-averaging (CA)-CFAR is considered optimal algorithm for exponentially distributed clutter [18]. However, due to the presence of outliers such as sidelobe contamination caused by target spectral leakage, the reference cells often fail to satisfy the homogeneity assumption, resulting in severe degradation of CA-CFAR performance. The CA greatest-of (CAGO)-CFAR [19] and the CA smallest-of (CASO)-CFAR [20], as two variants of the CA-CFAR, provide a more comprehensive consideration of extreme clutter conditions within the reference window. However, the performance of CAGO-CFAR degrades significantly under target masking effects, while CASO-CFAR tends to generate higher false alarm rates near clutter edges. The variability index (VI)-CFAR algorithm [21] leverages the advantages of mean-based CFAR algorithms. By analyzing the statistical dispersion of reference samples to compute the VI, it adaptively selects a suitable detection thresholding strategy. However, when the targets are located in heterogeneous environments or the background noise exhibits

Manuscript received November 1, 2025; accepted xxx xx, 2025. Date of publication xxx xx, 2025; date of current version xxx xx, 2025. This work was supported in part by the xxx (*Corresponding author: Fangjiong Chen.*)
Jiachen Zhu, and Fangjiong Chen are with the School of Electronic and Information Engineering, South China University of Technology, Guangzhou 510641, China. and also with the Guangdong Provincial Key Laboratory of Short-Range Wireless Detection and Communication, South China University of Technology, Guangzhou 510641, China. (e-mail: eezhujc@mail.scut.edu.cn; eefjchen@scut.edu.cn).
Jie Wu is with the Southern Marine Science and Engineering Guangdong Laboratory (Guangzhou), Guangzhou, 511458, China. (e-mail: wu_jie@gmlab.ac.cn).
Ming Xia is with the School of Electronics and Information Engineering, Beihang University, Beijing 100083, China. (e-mail: xiaming@buaa.edu.cn).
Digital Object Identifier

complex statistical characteristics, the spatial subsetting mechanism of VI-CFAR may fail to guarantee reliable detection performance.

To mitigate the adverse effects of interfering targets and clutter edges on threshold estimation, an effective approach is to sort and examine the reference samples [22]. Specifically, the ordered statistics (OS)-CFAR [23] arranges the reference cells in descending order and selects the $k$-th ranked sample as the background noise estimate for accurate threshold computation. However, this algorithm typically requires prior knowledge of the number of interfering targets or clutter edges to determine the optimal $k$ parameter setting. The trimmed mean (TM)-CFAR [24], a generalized extension of OS-CFAR, discards a certain number of the highest and lowest ranked samples and averages the remaining ones to estimate the background noise level. Studies have shown that the detection performance of TM-CFAR is susceptible to the trimming ratio, and its optimal parameters similarly depend on prior knowledge of the operating environment [25]. Furthermore, with the continuously improving resolution of millimeter-wave radar systems, the traditional sliding-window-based noise estimation strategies employed by conventional CFAR algorithms may no longer be effective [17].

For multi-target scenarios, the truncated statistics (TS)-CFAR [26] employs maximum likelihood estimation (MLE) to determine a truncated threshold, thereby achieving the separation of targets from anomalous interferences. However, when the number of observations is limited or the data distribution exhibits strong right-skewness, the MLE may become unstable or unsolvable. This limitation hinders the reliable determination of a threshold that satisfies the constraint conditions [27, 28]. Furthermore, in multi-target environments, abnormal interferences such as target sidelobes can contaminate the reference window samples, leading to biased noise estimation and degraded detection performance. To mitigate the influence of such outliers on threshold estimation, the CLEAN concept [29], which iteratively removes detected targets and their corresponding sidelobe artifacts, has been introduced into radar target detection.

The sparsity adaptive correlation maximization (SACM)-CFAR [30] achieves target detection by analyzing the correlation between the radar intermediate-frequency (IF) signal and the sensing matrix, and iteratively removes detected targets based on the CLEAN concept to effectively suppress sidelobe interference. Alternatively, the Sidelobe Suppression (SS)-CFAR separates target and noise components through second-order differential statistics and incorporates the CLEAN concept to mitigate target sidelobes. While these CLEAN-based target detection algorithms have demonstrated remarkable effectiveness in improving detection accuracy and suppressing sidelobe interference, certain limitations persist. Specifically, the SACM-CFAR relies on a compressive sensing-based detection framework, which makes maintaining a stable false alarm rate challenging under varying Signal-to-Noise Ratio (SNR) or sparsity conditions [31]. Meanwhile, the SS-CFAR employs a noise estimation strategy based on second-order differential statistics, which may introduce estimation bias, consequently degrading detection performance and increasing the false alarm probability. Therefore, the development of accurate background noise estimation methods and the design of a robust interference suppression mechanism are paramount to ensuring stable detection performance in complex environments.

The rapid evolution of deep learning (DL) has stimulated considerable interest in algorithms that exploit the multi-level feature abstraction capability of neural networks to extract radar echo characteristics [32]. The DL-CFAR [33], as the first CFAR detector based on neural networks, demonstrates superior performance over traditional approaches in mitigating masking effects and suppressing sidelobes. However, its training and validation datasets originate predominantly from anechoic chambers, which fail to account for the random noise and complex interference characteristic of real-world environments. Further advancing this field, the range-Doppler (RD) heatmap-based target detection approach [34] incorporates the spatial offset between the target and the local window center, and designs a detection head with both classification and regression functionalities to achieve target recognition and precise localization. Although such neural network-based detection algorithms provide new research directions for radar target detection, they still face challenges such as the difficulty of acquiring sufficient training data, high annotation costs, and limited model generalization. Therefore, further studies and improvements are required.

Based on the preceding analysis, this paper proposes a novel CFAR detection algorithm, the CLEAN and truncated-statistics CFAR (CT-CFAR), designed for sidelobe interference environments. The proposed algorithm employs TS to effectively separate target and noise components within the radar echo power spectrum, thereby restoring the homogeneity assumption of the reference window. Concurrently, we introduce learnable historical sidelobe information to enhance the robustness and environmental adaptability of the detection process. Furthermore, we develop a target reconstruction model by combining the Candan algorithm [35] with least-squares (LS) estimation [36] using multi-channel echo data, and incorporate the CLEAN concept [29], achieving effective suppression of target-induced sidelobe interference. The main contributions of this paper are summarized as follows:
1) TS-based noise modeling and estimation. We establish and analyze a mathematical model that considers the signal composition and statistical characteristics of the radar echo power spectrum after non-coherent accumulation (NCA). Based on the conditional expectation of background noise, we derive the relationship between TS and noise distribution, and design an iterative strategy to accurately estimate the echo noise, thereby restoring the homogeneity of the reference samples.
2) Development of an adaptive threshold estimation mechanism. We designed an adaptive threshold mechanism that incorporates a learnable historical sidelobe information matrix. The mechanism allows the detection threshold to dynamically adjust to the surrounding environment, thereby significantly enhancing the algorithm's robustness and environmental adaptability.
3) Construction of a CLEAN-based sidelobe suppression strategy. We construct a sidelobe suppression strategy based on the CLEAN concept. Specifically, we formulate a multi-channel target reconstruction model by integrating the Candan [35] algorithm with LS estimation using multi-channel echo data. Then employ the CLEAN concept to iteratively remove target components, which mitigates target-induced sidelobe interference and enhances detection accuracy.

The remainder of this article is organized as follows. Section II presents the problem formulation and its corresponding mathematical model. Section III details the proposed CT-CFAR algorithm. In Section IV, Monte Carlo simulations first validate the accuracy of CT-CFAR in background noise estimation, followed by evaluations of target detection performance and sidelobe suppression capability. Subsequently, experiments using real-world measurement data demonstrate the algorithm's effectiveness under practical conditions. Finally, Section V concludes the paper and discusses directions for future research.

## II. Problem Formulation

### A. Target Signal Model

An FMCW radar estimates the range and velocity of targets by transmitting a chirp signal whose instantaneous frequency increases linearly over time. The radar system generally comprises an antenna array, a frequency mixer, an analog-to-digital converter (ADC), and a digital signal processor [37]. Given the potential for nonlinearities in the transmitter, the transmitted waveform of the FMCW radar is defined as

$$S_\mathrm{T}(t) = A_\mathrm{T} e^{j\left[2\pi\left(f_c t + \frac{K}{2}t^2\right) + \phi_n\right]}, \ t \in [0,\ T_c] \qquad (1)$$

where $A_\mathrm{T}$ denotes the amplitude of the transmitted signal, $f_c$ denotes the initial carrier frequency of the chirp signal, $K = B/T_c$ denotes the sweep frequency slop, $B$ denotes the sweep bandwidth of the signal, $T_c$ denotes the period of the chirp signal, and $\phi_n$ denotes the phase noise.

When the electromagnetic wave propagates at the speed of light $c$, the received signal can be regarded as a time-delayed replica of the transmitted signal reflected from the target. At time $t$, for a target located at a distance $R_0$ from the radar and moving with a radial velocity $v$, the two-way propagation delay $\tau(t)$ can be defined as

$$\tau(t) = \frac{2(R_0 + vt)}{c}. \qquad (2)$$

Therefore, the received signal $S_\mathrm{R}(t)$ of an FMCW radar can be expressed as

$$S_\mathrm{R}(t) = A_\mathrm{R} e^{j\left[2\pi\left(f_c(t-\tau(t)) + \frac{K}{2}(t-\tau(t))^2\right) + \phi_n\right]} + n(t),\ t \in [0,\ T_c] \qquad (3)$$

where $A_\mathrm{R} = kA_\mathrm{T}$ denotes the amplitude of the received signal, $k$ denotes the channel loss, and $n(t)$ denotes the complex Gaussian additive noise.

By mixing the transmitted signal $S_\mathrm{T}(t)$ and the received signal $S_\mathrm{R}(t)$, and subsequently passing the result through a low-pass filter, the IF signal $S_\mathrm{IF}(t)$ can be obtained, which can be expressed as

$$S_\mathrm{IF}(t) = A_\mathrm{IF} e^{j2\pi(f_b + f_d)t + j\phi_0}, \qquad (4)$$

where $f_b = 2KR_0/c$ denotes the beat frequency, $f_d = 2vf_c/c$ denotes the Doppler frequency, $\phi_0$ denotes the phase constant.

To further discretize the IF echo model and extend it to the multi-target case, we assume that the radar receiver consists of a uniform linear array with $L$ antenna elements. The received signal is the superposition of echoes from $K$ targets. Each radar frame comprises $M$ chirps, and each chirp contains $N$ sampled points. The elements of the RD matrix $\mathbf{S}_l \in \mathbb{C}^{N \times M}$ corresponding to the $l$-th antenna can be expressed as

$$\mathbf{S}_l[n,m] = \sum_{k=1}^{K} A_{m,k} e^{j2\pi\left((f_{b,k} + f_{d,k})nT_s + f_{d,k}mT_{PRI}\right) + j\phi_0,k} a_l(\theta_k) + \mathbf{w}_l[n,m], \qquad (5)$$

where $A_{m,k}$, $f_{b,k}$, and $f_{d,k}$ denote the complex envelope, beat frequency, and Doppler frequency, respectively, of the $k$-th target IF signal. $a_l(\theta_k)$ denotes the $l$-th antenna of the steering vector of the $k$-th target, which is expressed as

$$a_l(\theta_k) = e^{-j2\pi \frac{(l-1)d\sin\theta_k}{\lambda}}, \qquad (6)$$

where $d$ denotes the isolation between antenna, $\theta_k$ denotes the incident angle of the $k$-th target, and $\lambda = c/f_c$ denotes the wavelength. Additionally, $\mathbf{w}_l[n,m] \sim \mathcal{CN}(0,\sigma^2)$ denotes the complex Gaussian white noise with zero mean and variance $\sigma^2$.

The multi-antenna received signal $\mathbf{S}_l$ is processed via two-dimensional fast Fourier transform (2D-FFT) along the fast-time and slow-time dimensions, yielding the power spectrum matrix that contains target information, also known as RD information (RDI). Subsequently, NCA is performed on the RDI along the antenna dimension to construct the CFAR detection matrix used for target detection. The overall signal processing flow for the millimeter-wave radar is shown in Fig. 1.

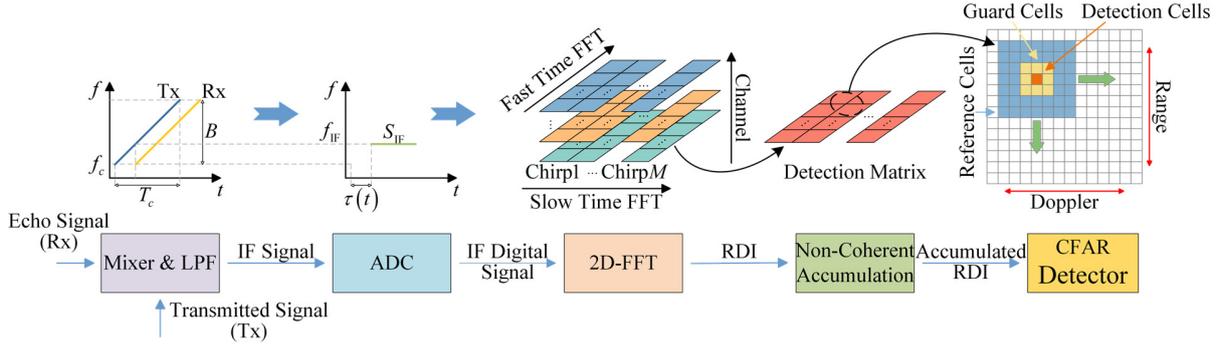

**Fig. 1.** FMCW radar signal processing flow. In the figure, $B$ is the sweep bandwidth, $f_c$ is the starting frequency, $T_c$ is the chirp duration, $\tau(t)$ is the echo time delay, and $f_{IF}$ is the intermediate frequency of the target echo.

### B. Hypothesis Testing for Target Detection

The CFAR detector of the FMCW radar estimates the background noise of the targets through a sliding window and sequentially detects each cell under test (CUT) in the RDI based on the locally estimated threshold to distinguish the target signal from the noise. To mitigate the influence of adjacent targets on the CUT, guard cells are typically set around the CUT. The corresponding target detection discriminant can be expressed as

$$x_{\text{CUT}}^{(r_{\text{Ind}}, v_{\text{Ind}})} \underset{\mathcal{H}_0}{\overset{\mathcal{H}_1}{\gtrless}} T, \tag{7}$$

where $x_{\text{CUT}}^{(r_{\text{Ind}}, v_{\text{Ind}})} \in \mathbb{R}^+$ denotes the CUT element located at the $r_{\text{Ind}}$-th row and $v_{\text{Ind}}$-th column of the CFAR detection matrix, $\mathcal{H}_0$ denotes the case where the CUT contains only noise, $\mathcal{H}_1$ denotes the case where the CUT contains both target and noise, $T = \alpha g(x_1, x_2, \ldots, x_N)$ denotes the local detection threshold corresponding to CUT, and $\{x_i\}_{i=1}^N$ denotes the reference cells in the sliding window. The nonlinear function $g$ maps $\{x_i\}_{i=1}^N$ to a non-negative real number, which serves as the local noise estimate of the current CUT. The scaling factor $\alpha$ is used to control the algorithm's desired false alarm probability $P_{\text{FA}}$.

## III. PROPOSED CT-CFAR DETECTOR

The CLEAN algorithm was originally proposed for radio interferometric imaging [29], where astronomers employed its iterative update mechanism to deconvolve the instrument's point spread function and reconstruct the true sky brightness distribution. Inspired by this concept, we introduce the CLEAN concept into FMCW radar target detection to mitigate the influence of target-induced sidelobes. Meanwhile, considering that the target energy is primarily concentrated in the main lobe and typically much higher than the background noise, we incorporate TS concept [26] to separate target and noise components, thereby restoring the homogeneity of reference cells and enabling mean-based threshold estimation for optimal detection performance. Therefore, this section proposes a robust CFAR target detection algorithm, the CT-CFAR, which achieves accurate detection in environments contaminated by sidelobes.

The flowchart of the CT-CFAR algorithm is shown in Fig. 2, which mainly consists of three components.

### A. Background Noise Estimation Based on TS

By applying a 2D-FFT to the multi-channel FMCW radar echoes as defined in (5), the power spectrum of the $l$-th channel, denoted as $\boldsymbol{S}_l' \in \mathbb{C}^{N \times M}$, is defined as

$$\begin{aligned}\boldsymbol{S}_l'[p,q] &= \sum_{n=0}^{N-1}\sum_{m=0}^{M-1} \boldsymbol{S}_l[n,m] e^{-j2\pi\frac{pn}{N}} e^{-j2\pi\frac{qm}{M}} \\ &= \sum_{k=1}^{K} e^{j\phi_{0,k}} e^{-j2\pi\frac{(l-1)d\sin\theta_k}{\lambda}} \boldsymbol{S}_n^{(k)}[p] \boldsymbol{S}_m^{(k)}[q] + \boldsymbol{W}_l[p,q] \\ &= \boldsymbol{X}_l[p,q] + \boldsymbol{W}_l[p,q], \end{aligned} \tag{8}$$

where the $\boldsymbol{S}_n^{(k)} \in \mathbb{C}^{N \times 1}$, $\boldsymbol{S}_m^{(k)} \in \mathbb{C}^{1 \times M}$, target information $\boldsymbol{X}_l \in \mathbb{C}^{N \times M}$, and noise information $\boldsymbol{W}_l \in \mathbb{C}^{N \times M}$, are defined as

$$\boldsymbol{S}_n^{(k)}[p] = \sum_{n=0}^{N-1} e^{j2\pi n\left((f_b + f_{d,k})T_s - \frac{p}{N}\right)}, \tag{9}$$

$$\boldsymbol{S}_m^{(k)}[q] = \sum_{m=0}^{M-1} A_{m,k} e^{j2\pi m\left(f_{d,k}T_{\text{PRI}} - \frac{q}{M}\right)}, \tag{10}$$

$$\boldsymbol{X}_l[p,q] = \sum_{k=1}^{K} e^{j\phi_{0,k}} e^{-j2\pi\frac{(l-1)d\sin\theta_k}{\lambda}} \boldsymbol{S}_n^{(k)}[p] \boldsymbol{S}_m^{(k)}[q], \tag{11}$$

$$\boldsymbol{W}_l[p,q] = \sum_{n=0}^{N-1}\sum_{m=0}^{M-1} w_l[n,m] e^{-j2\pi\frac{pn}{N}} e^{-j2\pi\frac{qm}{M}}. \tag{12}$$

Among them, $\boldsymbol{W}_l$ is assumed to follow a complex Gaussian distribution, defined as

$$\boldsymbol{W}_l[p,q] \sim \mathcal{CN}(0, NM\sigma^2). \tag{13}$$

To improve the detection probability, NCA is commonly employed to enhance the target signal power. After NCA across the $L$ channel, the CFAR detection matrix $\boldsymbol{P}_{\text{NCA}} \in \mathbb{R}^{N \times M}$ is defined as

$$\begin{aligned}\boldsymbol{P}_{\text{NCA}}[p,q] &= \sum_{l=1}^{L} |\boldsymbol{S}_l'[p,q]|^2 \\ &= \sum_{l=1}^{L} |\boldsymbol{X}_l[p,q]|^2 + \sum_{l=1}^{L} |\boldsymbol{W}_l[p,q]|^2 \\ &\quad + 2\sum_{l=1}^{L} \mathbb{R}\left(\boldsymbol{X}_l[p,q]\boldsymbol{W}_l^*[p,q]\right) \\ &= \boldsymbol{P}_{\text{s}}[p,q] + \boldsymbol{P}_{\text{n}}[p,q] + \boldsymbol{P}_{\text{sn}}[p,q], \end{aligned} \tag{14}$$

where $\mathbb{R}(\cdot)$ denotes the operation of taking the real part, the components $\boldsymbol{P}_{\text{s}} \in \mathbb{R}^{N \times M}$, $\boldsymbol{P}_{\text{n}} \in \mathbb{R}^{N \times M}$, and $\boldsymbol{P}_{\text{sn}} \in \mathbb{R}^{N \times M}$ correspond to the target component, the noise component, and the signal-noise cross-term component, respectively. These elements are explicitly defined as

$$P_s[p,q] = \sum_{l=1}^{L}|X_l[p,q]|^2, \quad (15)$$

$$P_n[p,q] = \sum_{l=1}^{L}|W_l[p,q]|^2, \quad (16)$$

$$P_{sn}[p,q] = 2\sum_{l=1}^{L}\mathbb{R}(X_l[p,q]W_l[p,q]). \quad (17)$$

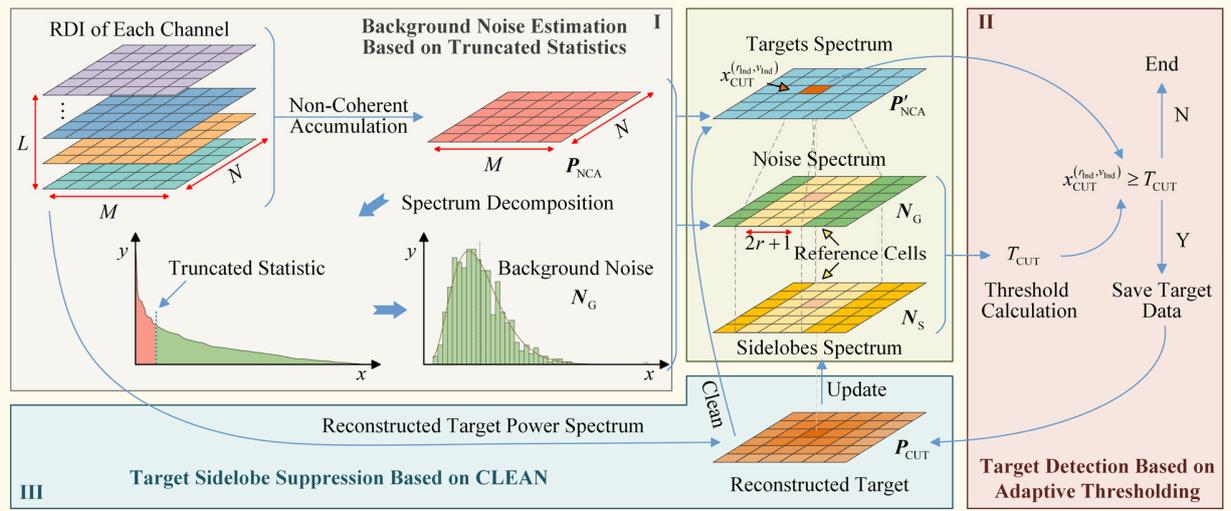

**Fig. 2.** Flowchart of the CT-CFAR algorithm.

For the component $P_n$, the complex noise term $W_l$ is modeled as a circularly symmetric complex Gaussian random variable, i.e., $W_l[p,q] \sim \mathcal{CN}(0, NM\sigma^2)$. Based on this assumption, the power of each complex sample follows an exponential distribution. Therefore, when summing over $L$ independent samples, $P_n$ follows a Gamma distribution with shape parameter $L$ and scale parameter $NM\sigma^2$, specifically

$$P_n[p,q] \sim \Gamma(L,\theta), \quad (18)$$

where $\theta = NM\sigma^2$.

In this case, the expectation $\mu_z$ of $P_n$ is readily obtained as

$$\mathbb{E}\left[\overline{P}_n\right] = L\theta = \mu_z \Rightarrow \theta = \mu_z/L, \quad (19)$$

where $\overline{P}_{(\cdot)}$ denotes the spatial average power of matrix $P_{(\cdot)}$, defined as

$$\overline{P}_{(\cdot)} = \frac{1}{NM}\sum_{q=1}^{N}\sum_{p=1}^{M}P_{(\cdot)}[p,q]. \quad (20)$$

Given that $L$ is known, the distribution of the background noise $P_n$ can thus be uniquely determined when $\mu_z$ is fixed.

To accurately estimate $P_n$ and thus achieve effective separation of target and noise components, we derive and present the complete solution process for $\mu_s$, starting from the conditional expectation of $P_n$. The details are presented as follows.

Since $W_l$ follows a circularly symmetric complex Gaussian distribution with zero mean, the expected value of $\overline{P}_{sn}$ is 0, which implies that

$$\mathbb{E}\left[\overline{P}_{NCA}\right] = \mathbb{E}\left[\overline{P}_s\right] + \mathbb{E}\left[\overline{P}_n\right], \quad (21)$$

In radar echoes, the target power is typically concentrated in the main lobe and is significantly higher than the background noise. The cross-term represents a zero-mean random fluctuation that diminishes as the number of NCA channels increases. Under medium-to-high SNR conditions, its statistical contribution to the total power spectrum is negligible. therefore, $P'_{NCA}$ can be regarded as a purely signal component containing only the target and its sidelobe information.

Based on this, when vectorizing $P_{NCA}$ and applying a truncation threshold $t$, the remaining samples $\{\tilde{x}_i\}_{i=1}^{N'}$ mainly originate from noise. Consequently, the mean of these truncated samples can be approximated by the conditional expectation of the noise power $P_n$ below the same threshold, which can be formalized as

$$\frac{1}{N'}\sum_{i=1}^{N'}\tilde{x}_i \approx \mathbb{E}(Z\mid Z \leq t), \quad (22)$$

where $Z$ denotes the noise power random variable $P_n$.

The PDF of $P_n$, $f_Z(z;\alpha,\beta)$, can be expressed as

$$f_Z(z;\alpha,\beta)\Big|_{\alpha=L,\beta=\frac{\mu_z}{L}} = \frac{1}{\beta^\alpha \Gamma(\alpha)}z^{\alpha-1}e^{-\frac{z}{\beta}}$$
$$= \frac{1}{(\mu_z/L)^L \Gamma(L)}z^{L-1}e^{-\frac{Lz}{\mu_z}}, \quad (23)$$

where $\alpha = L$ denotes the shape parameter of the Gamma distribution, and $\beta = \mu_z/L$ denotes the scale parameter of the Gamma distribution. The cumulative distribution function (CDF) of $P_n$, $F_Z(z)$, can be expressed as

$$F_Z(z;\alpha,\beta)\Big|_{\alpha=L,\beta=\frac{\mu_z}{L}} = \int_0^z f_Z(x;\alpha,\beta)dx = \frac{\gamma(L, Lz/\mu_z)}{\Gamma(L)}, \quad (24)$$

where $\gamma(\alpha,\beta) = \int_0^\beta x^{\alpha-1}e^{-x}dx$ denotes the incomplete Gamma function.

The conditional expectation of $P_n$ under the threshold $t$,

$\mathbb{E}(Z | Z \leq t)$, can be expressed as

$$\mathbb{E}(Z | Z \leq t) = \frac{\int_0^t x f_Z(x) dx}{F_Z(T)} = \mu_z \frac{\gamma(L+1, u)}{L\gamma(L, u)}, \quad (25)$$

where $u = Lt/\mu_z$.

Assuming that $P_n$ is truncated by the threshold $T$, the acceptable false alarm rate $P_{FA}$ satisfies $P(X > T) = P_{FA}$. Accordingly, the relationship between the CDF and $P_{FA}$ can be expressed as

$$1 - P_{FA} = F_Z(T) = \frac{\gamma\left(L, \frac{LT}{\mu_z}\right)}{\Gamma(L)} = \frac{\gamma(L, u_q)}{\Gamma(L)}, \quad (26)$$

where $u_q = u|_{t=T} = LT/\mu_z$.

As shown in (26), once $P_{FA}$ is given, $u_q$ is uniquely determined. The threshold $T$ can then be written as a function of $\mu_z$, $T(\mu_z) = u_q \mu_z / L$. Correspondingly, the conditional expectation $\mathbb{E}(Z | Z \leq t)$ can be further expressed as

$$\mathbb{E}(Z | Z \leq T(\mu_z)) = \mu_z g_u, \quad (27)$$

where $g_u = \gamma(L+1, u_q)/(L\gamma(L, u_q))$, which is a constant when $P_{FA}$ is given.

Based on the above relationship, once the acceptable false alarm probability $P_{FA}$ is given, and an initial estimate of $\mu'_z$ for the background noise distribution is obtained (e.g., by taking the median, mean, or truncated mean of the original sequence), the truncation threshold $T(\mu_z)$ is uniquely determined by $T(\mu_z) = u_q \mu_z / L$. With the truncated samples $\{\tilde{x}_i\}_{i=1}^{N'}$ thus obtained, the sample mean of these data satisfies

$$\frac{1}{N'}\sum_{i=1}^{N'} \tilde{x}_i \approx \mathbb{E}(Z | Z \leq T(\mu_z)) = \mu_z g_u. \quad (28)$$

Therefore, the mean value $\mu_z$ of the background noise $P_n$ can be estimated as

$$\hat{\mu}_z = \frac{1}{g_u N'} \sum_{i=1}^{N'} \tilde{x}_i. \quad (29)$$

The updated $\hat{\mu}_z$ is then substituted into the threshold equation $T(\mu_z)$ to obtain the updated truncation threshold $\hat{T}$. The samples are then truncated with this new threshold to further estimate the mean of the original distribution. These steps are iteratively repeated until $\hat{\mu}_z$ converges, and the convergence criterion is defined as

$$\frac{\hat{\mu}_z - \mu'_z}{\mu'_z + \varepsilon} < tol, \quad (30)$$

where $\mu'_z$ denotes the mean value of the previous estimate, $\varepsilon$ denotes the infinitesimally small positive regularization parameter introduced to ensure numerical stability, and *tol* denotes the tolerable error.

When the convergence criterion is satisfied, the estimated $\hat{\mu}_z$ represents the mean of the Gamma distribution corresponding to the background noise $P_n$. Accordingly, the scale parameter $\theta$ can be derived as $\theta = \hat{\mu}_z / L$, thus enabling accurate estimation of the background noise distribution in the power spectrum after the NCA of radar echoes. The estimated background noise matrix is denoted as $N_G \in \mathbb{R}^{N \times M}$, and the CFAR detection matrix is updated as

$$P'_{NCA} = P_{NCA} - N_G, \quad (31)$$

where $P'_{NCA} \in \mathbb{R}^{N \times M}$ denotes the updated CFAR detection matrix.

Since the fast-time and slow-time dimensions of FMCW radar are typically on the order of $10^2$ or higher, the resulting RD matrix contains tens of thousands of observation samples. Consequently, by applying the iterative algorithm described above, the mean of the truncated samples can effectively converge to the conditional expectation of the original (untruncated) Gamma distribution, thereby ensuring the consistency and unbiasedness of the estimation.

*B. Target Detection Based on Adaptive Thresholding*

In homogeneous environments, the CA-CFAR is considered the optimal detector model for exponentially distributed clutter [18]. In the previous subsection, we derived the relationship between the conditional expectation of the truncated statistic and the mean of the background noise, and achieved precise estimation of the echo background noise through iterative updates. The resulting estimated background noise matrix satisfies the assumptions of independence, identical distribution, and homogeneity, thereby fulfilling the prerequisite for mean-based CFAR detectors to achieve optimal detection performance.

On the other hand, to enhance the robustness of target detection, a learnable historical sidelobe accumulation matrix, $N_S \in \mathbb{R}^{N \times M}$, is introduced in the proposed CT-CFAR algorithm. This matrix serves as a dynamic, learned penalty factor to suppress false alarms induced by target sidelobe contamination.

$N_S$ is initialized to zero and iteratively learned throughout the detection process. Specifically, whenever a cell is confirmed as a target in the current iteration, a predefined neighborhood (e.g., a rectangular region of semi-width $r$) around the target cell is cumulatively updated within $N_S$. This update serves to quantify and record the predicted energy of the sidelobe interference emanating from that target. By incorporating $N_S$ into the threshold calculation, the deviation caused by missing sidelobe interference can be compensated for, thereby further improving the robustness and accuracy of the detection results.

On this basis, if the index of the CUT in $P'_{NCA}$ along the slow-time dimension is $v_{Ind}$, then its detection threshold can be defined as

$$T_{CUT} = \frac{\alpha}{N + 2r + 1} \sum_{i=1}^{N} \sum_{j=v_{Ind}-r}^{v_{Ind}+r} (N_G[i,j] + N_S[i,j]), \quad (32)$$

where $\alpha$ denotes the multiplication factor that controls the $P_{FA}$, and $r$ denotes the reference window half-length of the target in the slow time dimension. In our experimental setup, $r$ is set to 2, following established empirical practices [30, 38], to minimize the extent of the local region while ensuring adequate coverage of the target's main lobe.

*C. Target SS Based on CLEAN*

Assuming the FMCW radar possesses $L$ array channels and the CUT is the largest target value in $\boldsymbol{P}'_{\mathrm{NCA}}$, it can be expressed as

$$x_{\mathrm{CUT}}^{(r_{\mathrm{Ind}}, v_{\mathrm{Ind}})} = \max\left(\boldsymbol{P}'_{\mathrm{NCA}}\right). \tag{33}$$

The presence of a target at $\boldsymbol{P}'_{\mathrm{NCA}}[r_{\mathrm{Ind}}, v_{\mathrm{Ind}}]$ is determined by the condition $x_{\mathrm{CUT}}^{(r_{\mathrm{Ind}}, v_{\mathrm{Ind}})} > T_{\mathrm{CUT}}$, in which case the target information is recorded.

To achieve effective target SS, the CFAR detection matrix $\boldsymbol{P}'_{\mathrm{NCA}}$ can be subtracted by the non-coherently accumulated power spectrum of the current target, which effectively removes the target information contained in $\boldsymbol{P}'_{\mathrm{NCA}}$. However, since the detected frequencies are aligned with the discrete sampling grid, reconstructing the target echo according to the detected beat and Doppler frequencies using (4) will result in a peak-like characteristic at the corresponding target position in the processed signal's spectrum. To address this issue, the Candan frequency refinement algorithm [35] is employed to obtain high-precision estimates of the target's beat and Doppler frequencies, thereby improving the accuracy of target echo reconstruction.

A complex exponential signal $y(n)$ is defined as

$$y(n) = A e^{j(2\pi f n + \phi)}, \; n = [0, 1, \ldots, N-1] \tag{34}$$

where $f$ denotes the signal frequency, which can be represented in terms of the FFT frequency bin as $f = (n_p + \delta)/N$. Here, $n_p$ denotes the integer frequency bin index, and $\delta$ denotes the positive real number located within $[-1/2, 1/2]$ [39].

The Candan algorithm calculates the true frequency of the target based on the complex values of three adjacent points: $y(n_p - 1)$, $y(n_p)$, and $y(n_p + 1)$. This is achieved by estimating the deviation $\delta$ between the true position of the target peak and the index $n_p$. The deviation $\delta$ is defined as

$$\delta = C_N \, \mathbb{R}\left( \frac{y(n_p+1) - y(n_p-1)}{2y(n_p) - y(n_p+1) - y(n_p-1)} \right), \tag{35}$$

where $y(n_p)$ represents the complex value of the frequency-bin sample at $n_p$, and $C_N = \{1, 2, \ldots\}$ is the window parameter.

For the power spectrum matrix $\boldsymbol{S}_l$ of each channel, three complex samples indexed by $[r_{\mathrm{Ind}}-1, v_{\mathrm{Ind}}]$, $[r_{\mathrm{Ind}}, v_{\mathrm{Ind}}]$, and $[r_{\mathrm{Ind}}+1, v_{\mathrm{Ind}}]$ are successively extracted along the fast-time dimension. The fractional frequency bias of the detected target in each channel, $\{\delta_{Rl}\}_{l=1}^{L}$, is then calculated using (35). Similarly, three complex samples indexed by $[r_{\mathrm{Ind}}, v_{\mathrm{Ind}}-1]$, $[r_{\mathrm{Ind}}, v_{\mathrm{Ind}}]$, and $[r_{\mathrm{Ind}}, v_{\mathrm{Ind}}+1]$ are successively extracted along the slow-time dimension from the power spectrum matrix $\boldsymbol{S}_l$ of each channel. The fractional frequency bias of the detected target in the Doppler frequency for each channel, $\{\delta_{Vl}\}_{l=1}^{L}$, is then calculated using (35).

By averaging the frequency biases $\{\delta_{Rl}\}_{l=1}^{L}$ and $\{\delta_{Vl}\}_{l=1}^{L}$ separately, the high-precision range index $\hat{r}_{\mathrm{Ind}}$ and velocity index $\hat{v}_{\mathrm{Ind}}$ corresponding to $x_{\mathrm{CUT}}^{(r_{\mathrm{Ind}}, v_{\mathrm{Ind}})}$ can be expressed as

$$\hat{r}_{\mathrm{Ind}} = r_{\mathrm{Ind}} + \frac{1}{L}\sum_{l=1}^{L}\delta_{Rl}, \tag{36}$$

$$\hat{v}_{\mathrm{Ind}} = v_{\mathrm{Ind}} + \frac{1}{L}\sum_{l=1}^{L}\delta_{Vl}, \tag{37}$$

where $\hat{r}_{\mathrm{Ind}}$ and $\hat{v}_{\mathrm{Ind}}$ denote the high-precision beat frequency and Doppler frequency indices corresponding to $x_{\mathrm{CUT}}^{(r_{\mathrm{Ind}}, v_{\mathrm{Ind}})}$, respectively. Based on the range and velocity resolution of the FMCW radar, the target's distance and velocity can then be accurately estimated.

After accurately estimating the target's distance and velocity, substituting these into the $\boldsymbol{X}_l$ term of (8) yields the power spectrum, $\boldsymbol{G}_l \in \mathbb{C}^{N \times M}$, corresponding to the single-channel echo after 2D-FFT. However, the amplitude of this power spectrum is still unknown, necessitating further estimation of its complex envelope to achieve precise characterization of the target signal. To effectively address the issue of potential gain and phase inconsistencies among different antennas, we utilize multi-antenna data within the target region of the RD matrix to construct a LS-based multi-antenna fitting method, thereby enabling robust reconstruction of the target's non-coherently accumulated power spectrum distribution. The specific algorithmic process is as follows.

Considering that the target signal is typically concentrated in the main lobe and adjacent sidelobes, we truncate a neighborhood centered at $\boldsymbol{G}_l[r_{\mathrm{Ind}}, v_{\mathrm{Ind}}]$ to serve as the template matrix, which is expressed as

$$\boldsymbol{G}_p \in \mathbb{C}^{r \times r}. \tag{38}$$

The template matrix is vectorized column by column into a vector $\boldsymbol{g} \in \mathbb{C}^{2r \times 1}$.

The observation data $\{\boldsymbol{S}_l\}_{l=1}^{L}$ are processed by extracting the corresponding neighborhood based on the template matrix $\boldsymbol{G}_p$, resulting in the observation tensor $\boldsymbol{\mathcal{Y}} \in \mathbb{C}^{r \times r \times L}$. By unfolding this tensor channel by channel, we obtain the observation matrix $\boldsymbol{Y} \in \mathbb{C}^{2r \times L}$, which is expressed as

$$\boldsymbol{Y} = [\boldsymbol{y}_1, \boldsymbol{y}_2, \ldots, \boldsymbol{y}_L], \tag{39}$$

where $\boldsymbol{y}_l \in \mathbb{C}^{2r \times 1}$ denotes the observation vector for the $l$-th channel.

The relationship between the observation matrix $\boldsymbol{Y}$ and the template vector $\boldsymbol{g}$ can be modeled as

$$\boldsymbol{Y} = \boldsymbol{g}\boldsymbol{\alpha} + \boldsymbol{n}, \tag{40}$$

where $\boldsymbol{\alpha} = [\alpha_1, \alpha_2, \ldots, \alpha_L] \in \mathbb{C}^{1 \times L}$ denotes the channel complex gain vector, $\alpha_l$ denotes the complex coefficient of the $l$-th antenna, and $\boldsymbol{n} \in \mathbb{C}^{2r \times L}$ denotes the additive noise matrix.

Based on the LS criterion, the optimal complex gain vector $\boldsymbol{\alpha}$ can be obtained as

$$\hat{\boldsymbol{\alpha}} = \arg\min_{\boldsymbol{\alpha}} \|\boldsymbol{Y} - \boldsymbol{g}\boldsymbol{\alpha}\|_{\mathrm{F}}^{2}, \tag{41}$$

where $\|\cdot\|_{\mathrm{F}}$ denotes the Frobenius norm. Consequently, the objective function $J(\boldsymbol{\alpha})$ is defined as

**Algorithm 1** CT-CFAR

**Input:** $S_l$, $P_{FA}$, $tol$
**Initialize:** $P_{NCA} \leftarrow \sum_{l=1}^{L} |S_l|^2$, $N_S \leftarrow 0$
$N_G$, $\hat{T} \leftarrow$ Noise estimation$(P_{NCA}, P_{FA}, tol)$
$P'_{NCA} \leftarrow P_{NCA} - N_G$
**While** $P'_{NCA}$ is not empty:
$\quad x_{CUT}^{(r_{Ind}, v_{Ind})} \leftarrow \max(P'_{NCA})$, $T_{CUT} \leftarrow$ Threshold$(N_G, N_S)$
$\quad$ **If** $x_{CUT}^{(r_{Ind}, v_{Ind})} > T_{CUT}$
$\quad\quad$ Save the target information
$\quad\quad P_{CUT} \leftarrow$ Reconstruct power spectrum$(S_l, x_{CUT}^{(r_{Ind}, v_{Ind})})$
$\quad\quad P'_{CUT} \leftarrow P_{CUT}(r_{Ind}, v_{Ind}) = 0$
$\quad\quad P'_{NCA} \leftarrow P'_{NCA} - P_{CUT}$, $N_S \leftarrow N_S + P'_{CUT}$
$\quad$ **Else**
$\quad\quad$ Break
**Output:** Detection result, $\hat{T}$

$$J(\alpha) = \|Y - g\alpha\|_F^2 = \text{tr}\left((Y - g\alpha)^H (Y - g\alpha)\right), \quad (42)$$

where $\text{tr}(\cdot)$ denotes the trace of the matrix. $J(\alpha)$ can be further expanded as

$$J(\alpha) = \text{tr}(Y^H Y - Y^H g\alpha - \alpha^H g^H Y + \alpha^H g^H g\alpha). \quad (43)$$

By taking the derivative of $J(\alpha)$ with respect to $\alpha^H$ and setting it to zero, the closed-form solution for $\hat{\alpha} \in \mathbb{C}^{1 \times L}$ is obtained as

$$\hat{\alpha} = \frac{g^H Y}{g^H g + \varepsilon}, \quad (44)$$

where $\varepsilon$ denotes a very small positive regularization parameter, used to ensure the stability of numerical computation. Consequently, the estimated power spectrum vector $\hat{g} \in \mathbb{C}^{2r \times L}$ can be expressed as

$$\hat{g} = g\hat{\alpha}. \quad (45)$$

This estimated vector $\hat{g}$ is then reshaped into a three-dimensional tensor $\hat{G} \in \mathbb{C}^{r \times r \times L}$ and subsequently cropped to $\hat{G}' \in \mathbb{C}^{N \times M \times L}$ via zero-padding. The resulting power spectrum matrix after NCA along the channel dimension is denoted by $P_{CUT} \in \mathbb{R}^{N \times M}$. Consequently, based on the CLEAN concept, the CFAR detection matrix $P'_{NCA}$ can be updated as

$$P'_{NCA} = P'_{NCA} - P_{CUT}. \quad (46)$$

Simultaneously, the learnable historical sidelobe matrix, $N_S$, is updated as

$$N_S = N_S + P'_{CUT}, \quad (47)$$

where $P'_{CUT}$ denotes the matrix that retains only the target sidelobe information after setting $P_{CUT}[r_{Ind}, v_{Ind}]$ to zeros.

Through the above steps, effective suppression of target sidelobes can be achieved. This iterative process continues until the CFAR detector no longer identifies any cells in the updated matrix $P'_{NCA}$ that satisfy the detection threshold, at which point the target estimation of the radar echo is considered complete.

The CT-CFAR algorithm is summarized in Algorithm 1.

## IV. EXPERIMENTS

In this section, radar echoes are first simulated under an additive complex Gaussian noise environment to evaluate the performance metrics of the proposed target detector. Subsequently, the algorithm is further validated using real human posture data collected by a cascaded millimeter-wave radar. Finally, the execution efficiency is analyzed based on the runtime of each algorithm. To ensure the reliability of the experimental results, all performance metrics are obtained through 100 independent standard Monte Carlo trials. In the simulation experiments with a given SNR, the noise power is set according to the average power of the noiseless simulated echo signal. The main parameters used in the FMCW radar simulation experiments are listed in TABLE 1.

TABLE 1
SIMULATION PARAMETERS

| Parameter | Value |
| --- | --- |
| Starting frequency | 77 GHz |
| Sweep slope | 120.023 MHz/us |
| Sweep bandwidth | 3.413 GHz |
| Sampling rate | 9000 Ksps |
| Number of samples per chirp | 256 |
| Number of samples per frame | 128 |

*A. Evaluation Metrics*

1) Probability of Detection

The detection performance of a target detection algorithm is typically characterized by the detection probability $P_d$, also referred to as the recall rate, which is defined as

$$P_d = \frac{N_{TP}}{N_P}, \quad (48)$$

where $N_{TP}$ denotes the number of true positive samples (correctly detected), and $N_P$ denotes the total number of positive samples. A higher recall rate signifies superior detection performance, reflecting the actual detection capability of the detector.

2) Probability of False Alarm

The false alarm probability $P_{fa}$ is an important metric for evaluating the false reporting rate of FMCW radar target detection algorithms [40], and is defined as

$$P_{fa} = \frac{N_{FP}}{N_N}, \quad (49)$$

where $N_{FP}$ denotes the number of false positive samples (negative samples incorrectly identified as positive), and $N_N$ denotes the total number of negative samples. A lower $P_{fa}$ indicates fewer false detections, corresponding to better detection performance. It should be noted that $P_{FA}$ is a preset hyperparameter that controls the false alarm probability of the target detector, and it differs from the empirically measured $P_{fa}$ defined here. In practical applications, $P_{fa}$ and $P_d$ are often mutually constrained, requiring a trade-off between detection sensitivity and false alarm control.

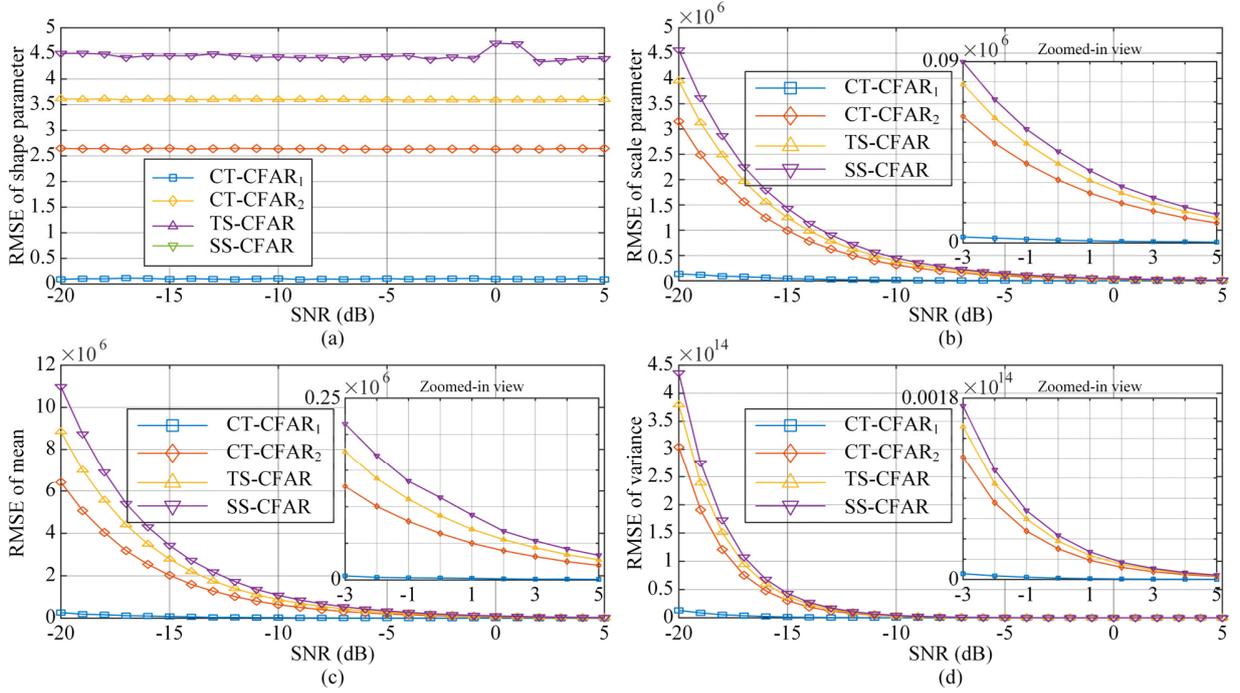

**Fig. 3.** Performance of Gamma parameter estimation methods across different SNRs, the inset shows an enlarged view of the SNR range from -3dB to 5dB. (a) RMSE of shape parameter. (b) RMSE of scale parameter. (c) RMSE of mean. (d) RMSE of variance.

3) Probability of Accurate Detection

The accuracy $P_a$ is an important indicator for evaluating the overall performance of a target detection algorithm. It intuitively reflects the global performance of the detector by jointly considering the correct detections of both positive and negative samples. It is defined as

$$P_a = \frac{N_{TP}}{N_{TP} + N_{FP}}. \quad (50)$$

*B. Background Noise Estimation Results*

According to (18), the background noise in the non-coherently accumulated echo power spectrum adheres to a Gamma distribution. To evaluate the performance of the proposed CT-CFAR in noise estimation, we employ the background noise estimation methods from SS-CFAR [38] and TS-CFAR [26] as comparative baselines. Specifically, the background noise directly estimated by CT-CFAR and that obtained based on the estimated truncation threshold are compared with those of the two baseline methods and the true background noise distribution, respectively. This set of comparisons is designed to assess the reconstruction accuracy of the noise component in the echo power spectrum for each algorithm.

The background noise estimation performance of various target detection algorithms is evaluated under different SNR conditions through simulations based on radar echoes that include 20 randomly located targets with randomized ranges and velocities. Fig. 3 illustrates the root-mean-square error (RMSE) between the estimated and true background noise parameters obtained by each algorithm as the SNR varies. In this figure, CT-CFAR$_1$ represents the noise distribution estimated by CT-CFAR, while CT-CFAR$_2$ denotes the noise distribution derived by applying the truncation threshold (estimated by CT-CFAR) to the observed samples. Experimental results show that as the SNR increases, the RMSE of the estimated noise distribution parameters relative to the original noise gradually decreases for all methods, and CT-CFAR$_1$ consistently maintains the optimal background noise estimation performance.

Fig. 4 shows the background noise estimation results of different target detection algorithms under an SNR=0dB. Subplots (a)-(e) illustrate the PDFs of the true noise, the noise directly estimated by CT-CFAR, the noise obtained by CT-CFAR with truncation thresholding, the noise estimated by SS-CFAR using second-order difference-based truncation statistics, and the noise estimated by TS-CFAR using threshold-based truncation, respectively. The red curves indicate the MLE fitted distributions. The experimental results show that the CT-CFAR$_1$ (Fig. 4 (b)) exhibits the highest morphological consistency with the true noise distribution (Fig. 4 (a)). Fig. 4 (f) further depicts the corresponding quantile-quantile (Q-Q) plot, where the data points of the CT-CFAR$_1$ align more closely with the theoretical quantiles (diagonal line) than those of the other methods, indicating the best fitting performance.

The robustness of each algorithm in background noise estimation is further examined at an SNR of 0dB. Fig. 5 presents boxplots showing the statistical characteristics of estimation errors for each method. The elliptical regions highlight the median and 95% confidence intervals of the estimation errors. Experimental results show that CT-CFAR$_1$ achieves the median closest to the true distribution and the narrowest confidence interval, indicating superior accuracy, stability, and consistency with the true noise distribution compared to the other detection algorithms.

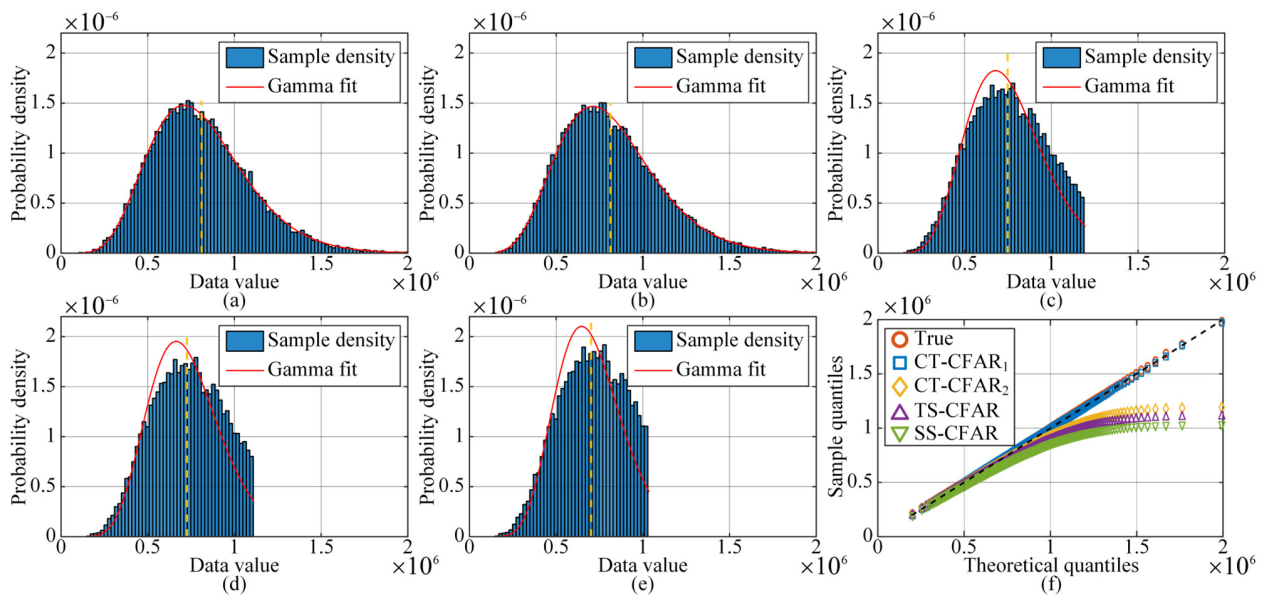

**Fig. 4.** Comparison of background noise modeling and distribution fitting results for different estimation methods at SNR = 0dB. (a) True background noise. (b) CT-CFAR$_1$ fitting. (c) CT-CFAR$_2$ fitting. (d) SS-CFAR fitting. (e) TS-CFAR fitting. (f) Q-Q plot of the five distributions.

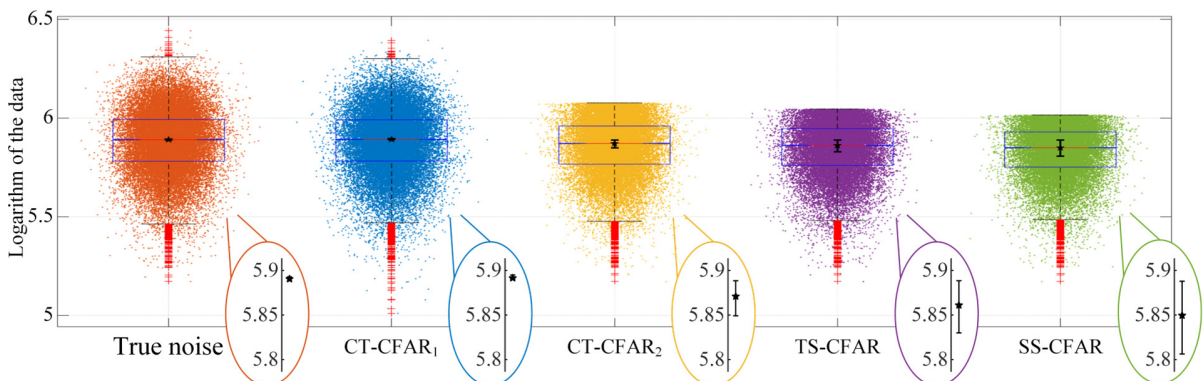

**Fig. 5.** Boxplot of algorithm results with median and 95% bootstrap confidence intervals at SNR = 0dB. The elliptical regions highlight the median and 95% confidence intervals of the estimation errors.

*C. Detection Performance Evaluation*

1) Simulation Experiments

A Monte Carlo simulation is conducted to comprehensively evaluate the target detection performance of the proposed CT-CFAR algorithm. In each independent trial, the simulated radar echoes contain 20 targets randomly distributed in range and velocity. To emulate realistic scenarios with stationary objects, up to half of the targets are randomly assigned zero velocity in each run. All performance metrics are averaged over 100 Monte Carlo trials to ensure statistical reliability and objectivity.

For fair comparison, the $P_{FA}$ of all CFAR detectors is uniformly set to $10^{-3}$. In sliding-window-based detectors, the numbers of half-window reference cells are set to 10 in the fast-time dimension and 6 in the slow-time dimension, with 5 guard cells on each side. The OS-CFAR employs the median as its order statistic, while the TM-CFAR uses three cut cells on each side for truncation. For the proposed CT-CFAR, the false-alarm probability for internal noise estimation $P_{FA}$ and the convergence tolerance $tol$ for the iterative process are set to $10^{-3}$ and $10^{-5}$, respectively.

Fig. 6 shows the detection performance of the proposed CT-CFAR algorithm with that of other target detection algorithms under different SNR conditions. Specifically, Fig. 6 (a) shows the detection probability $P_d$ of each algorithm as a function of the SNR. As illustrated, the $P_d$ of all algorithms increases monotonically with SNR and gradually saturates around -15dB, beyond which the improvement rate slows down. Among all the compared algorithms, CT-CFAR consistently exhibits the best detection performance across most SNR ranges. In the low-SNR region (SNR < -17dB), OS-CFAR and CASO-CFAR temporarily outperform others due to their tendency to adopt lower detection thresholds. However, as SNR increases, CT-CFAR and SS-CFAR surpass them and maintain superior detection probabilities. CASO-CFAR, TM-CFAR, and TS-CFAR also achieve competitive performance, but their detection capability is degraded by the influence of target sidelobes and outliers within the reference window, which bias the threshold estimation. CA-CFAR suffers from a similar issue because its reference cells fail to meet the background homogeneity assumption, leading to inferior detection performance. In addition, CAGO-CFAR shows the lowest detection sensitivity in multi-target scenarios owing to the masking effect. Overall, these results confirm that CT-CFAR achieves the highest detection sensitivity and robustness among all compared algorithms, demonstrating its effectiveness under varying SNR conditions.

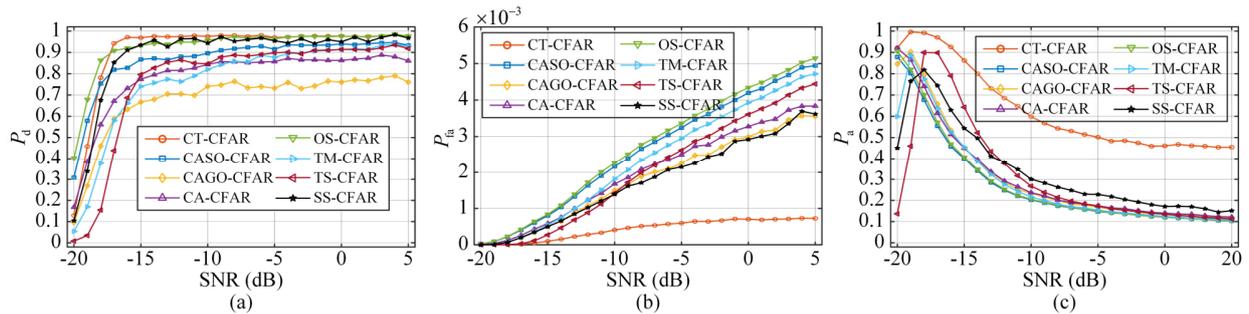

**Fig. 6.** Performance evaluation of different CFAR algorithms for various SNRs at $P_{FA} = 10^{-3}$. (a) Detection probability. (b) False alarm probability. (c) Precision.

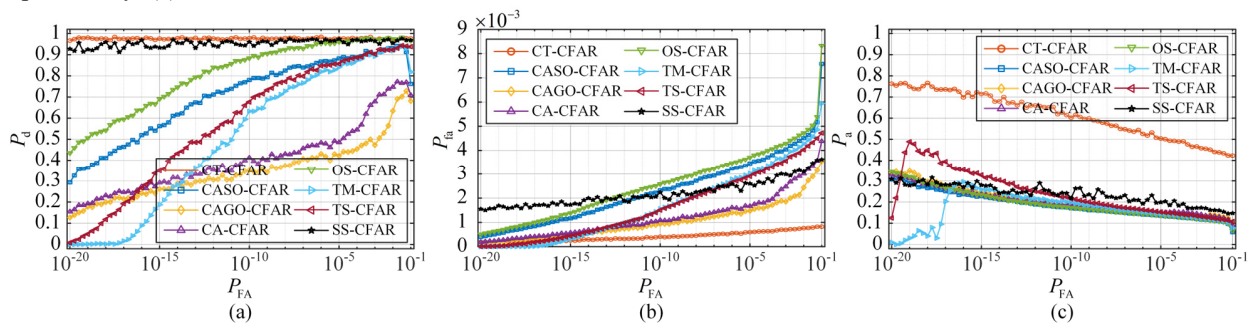

**Fig. 7.** Performance evaluation of different CFAR algorithms for various false alarm probability at SNR=0dB. (a) ROC curves. (b) False alarm probability. (c) Precision.

Fig. 6 (b) shows the false alarm probability $P_{fa}$ of each target detection algorithm as a function of SNR. As illustrated in the figure, CT-CFAR is the only algorithm that maintains the lowest and nearly constant false alarm probability across the entire SNR range. In contrast, the $P_{fa}$ of all other algorithms increases significantly with rising SNR. Among the others, OS-CFAR, CASO-CFAR, TM-CFAR, and TS-CFAR exhibit the most rapid increase in $P_{fa}$. This is mainly because a higher SNR introduces more target sidelobes and noise components that are mistakenly detected as targets, thereby increasing the number of false alarms. In comparison, CAGO-CFAR and CA-CFAR demonstrate stronger SS capability, but at the cost of reduced target detection performance (i.e., lower $P_d$). SS-CFAR maintains a relatively low false alarm rate while achieving detection performance comparable to CT-CFAR, making its overall performance second only to CT-CFAR.

Fig. 6 (c) shows the precision $P_a$ of each target detection algorithm as a function of SNR. As illustrated, although the overall precision of CT-CFAR slightly decreases with increasing SNR, it consistently maintains the highest level across the entire SNR range, demonstrating significantly superior performance compared to the other algorithms. In contrast, the precision of the remaining detection algorithms initially increases and then decreases, a trend consistent with the observations in Fig. 6 (b). Overall, CT-CFAR also exhibits the best performance in terms of precision, maintaining a low false alarm rate while achieving high target recall and precision, which proves its robustness and reliability in target detection.

Fig. 7 presents the detection performance of various algorithms at an SNR=0dB with respect to the $P_{FA}$. Fig. 7 (a) shows the receiver operating characteristic (ROC) curves for all algorithms. As observed, both CT-CFAR and SS-CFAR exhibit relatively small performance fluctuations with respect to the $P_{FA}$ and maintain stable detection results. Among them, the proposed CT-CFAR consistently achieves superior performance across the entire SNR range, demonstrating higher detection accuracy and robustness than all other algorithms.

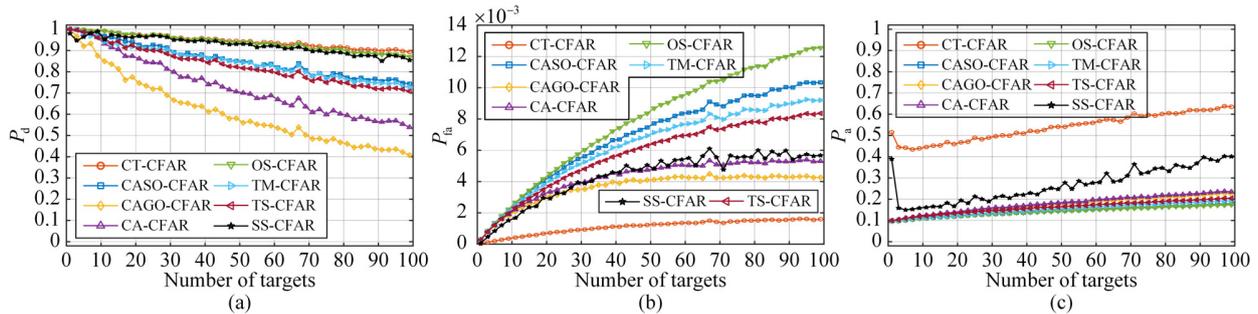

**Fig. 8.** Performance evaluation of different CFAR algorithms for various numbers of targets at $P_{FA}=10^{-3}$ and SNR=0dB. (a) Detection probability. (b) False alarm probability. (c) Precision.

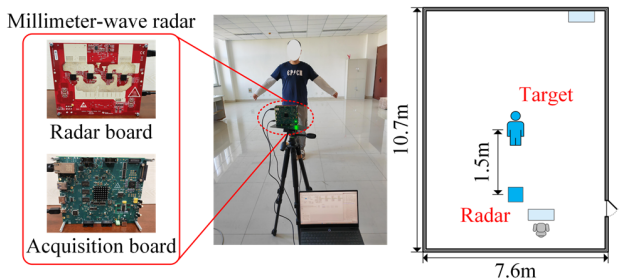

**Fig. 9.** Measurement environment. A subject stands with arms half-raised.

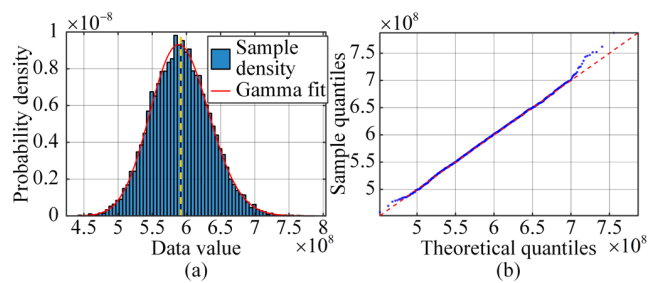

**Fig. 10.** Noise estimation performance of CT-CFAR in the measurement environment. (a) background noise distribution estimation. (b) Q-Q plot.

Fig. 7 (b) and Fig. 7 (c) respectively present the curves of the $P_{fa}$ and $P_a$ as a function of the prescribed $P_{FA}$. Consistent with the ROC curve results, both CT-CFAR and SS-CFAR exhibit strong stability with respect to variations in $P_{FA}$. However, CT-CFAR shows even smaller performance fluctuations and maintains more consistent metrics across different $P_{FA}$ levels, indicating superior robustness compared to SS-CFAR and the other algorithms.

Fig. 8 shows the performance comparison of each target detection algorithms in scenarios with varying numbers of targets under the conditions of $P_{FA}=10^{-3}$ and SNR=0dB. It can be observed that as the number of targets in the echo increases, the detection capability of all algorithms generally decreases. CT-CFAR consistently maintains the optimal detection performance across all target density scenarios. Nevertheless, in terms of $P_{fa}$ and $P_a$, CT-CFAR significantly outperforms the other algorithms. Overall, although CT-CFAR's detection capability is slightly inferior to the best-performing algorithms in high target density scenarios, it still maintains a high level of stability and detection performance.

2) Measurement Experiments

An experiment in a 10.7 m × 7.6 m empty classroom is carried out to assess the real-world detection performance of the proposed CT-CFAR algorithm. The subject, positioned 1.5m away from the radar, maintained a half-raised arm posture. Data are collected using a TI AWR2243 radar comprising a four-chip cascaded array [41], which is mounted on a tripod at a height of 1.2 m. Detailed radar parameters are listed in TABLE 2, and the experimental setup is shown in Fig. 9.

Fig. 10 shows the noise estimation performance of the CT-CFAR algorithm in the measurement environment. Fig. 10 (a) shows the estimated background noise and its corresponding fit. Since the cascaded radar possesses 192 array channels after MIMO virtualization, its distribution can be approximated by a Gaussian distribution according to the central limit theorem. In the Shapiro-Wilk test [42] conducted at a significance level of 0.01, the CT-CFAR estimated background noise failed to reject the Gaussian distribution hypothesis, the empirical fitting results are consistent with theoretical analysis. Fig. 10 (b) shows the corresponding Q-Q plot, confirming that the estimated Gamma distribution effectively approximates the statistical characteristics of the true background noise.

Fig. 11 shows the side-view projection of target detection results in the range dimension for CT-CFAR and several comparative CFAR algorithms. In the figure, the thick green line represents the zero-Doppler component, while red boxes indicate potential false alarm detections. Since only stationary targets are present in the measurement scenario, the true targets are theoretically expected to lie primarily along the

> REPLACE THIS LINE WITH YOUR MANUSCRIPT ID NUMBER (DOUBLE-CLICK HERE TO EDIT) <

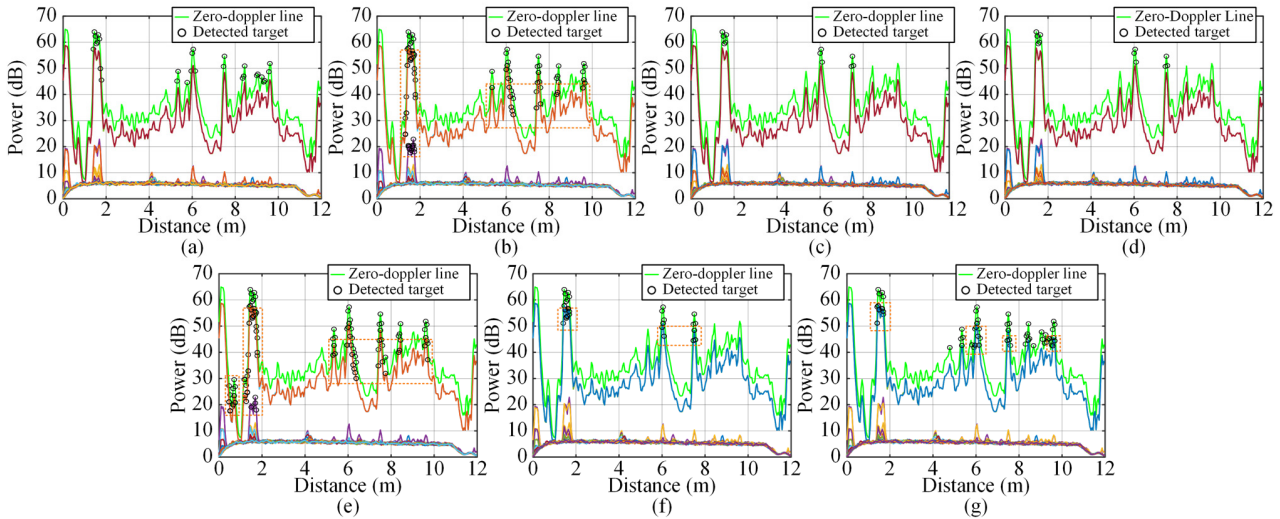

**Fig. 11.** Comparison of range-domain side-view projections of RDI detection results obtained by different CFAR algorithms. Red boxes denote potential false-alarm targets. (a) CT-CFAR. (b) CASO-CFAR. (c) CAGO-CFAR. (d) CA-CFAR. (e) OS-CFAR. (f) TM-CFAR. (g) SS-CFAR.

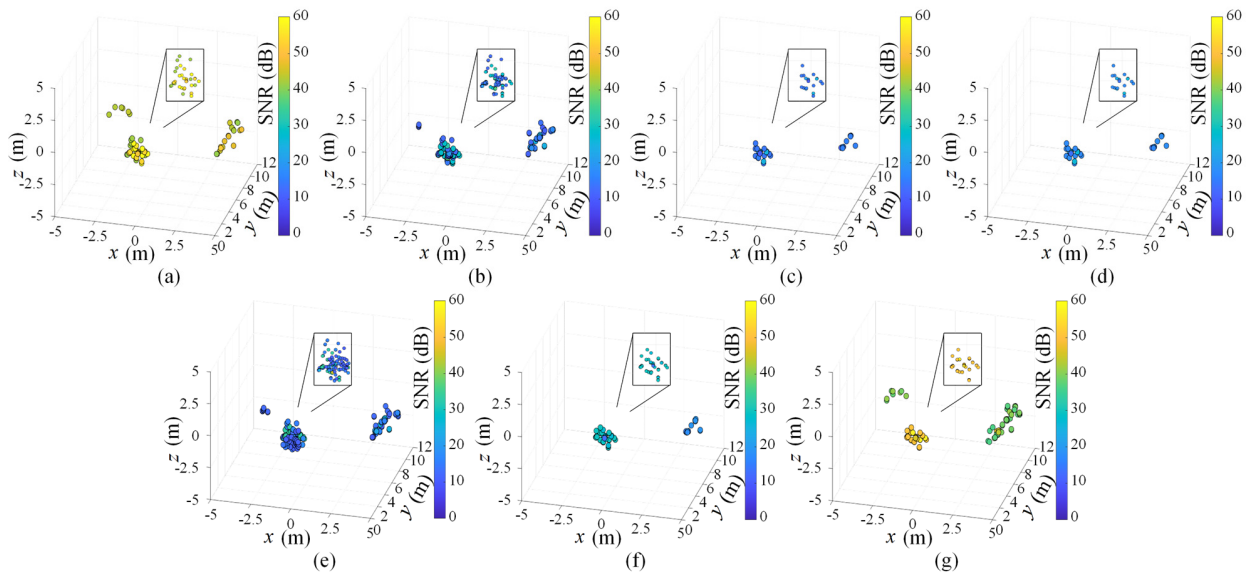

**Fig. 12.** Comparison of radar point clouds obtained by different CFAR target detection algorithms. The box shows an enlarged view of the human posture region. (a) CT-CFAR. (b) CASO-CFAR. (c) CAGO-CFAR. (d) CA-CFAR. (e) OS-CFAR. (f) TM-CFAR. (g) SS-CFAR.

zero-Doppler line.

TABLE 2
Radar Configuration Parameters

| Parameter | Value |
| --- | --- |
| Starting frequency | 77 GHz |
| Sweep slope | 100 MHz/us |
| Sweep bandwidth | 3.2 GHz |
| Sampling rate | 8000 Ksps |
| Number of samples per chirp | 256 |
| Number of samples per frame | 64 |

The experimental results indicate that CT-CFAR and SS-CFAR exhibit the best overall detection performance, effectively mitigating spectral leakage along both fast-time and slow-time dimensions. However, the target-noise separation strategy employed by SS-CFAR, which relies on second-order difference statistics, may lead to a biased threshold estimation, resulting in a marginally higher false alarm rate compared to CT-CFAR. CAGO-CFAR, CA-CFAR, and TM-CFAR also perform well in mitigating spectral leakage, but their detection sensitivity is relatively low due to the masking effect. In contrast, OS-CFAR and CASO-CFAR demonstrate strong detection capability for weak targets, yet their insufficient suppression of noise and sidelobe energy increases the likelihood of misclassification, leading to elevated false alarm rates.

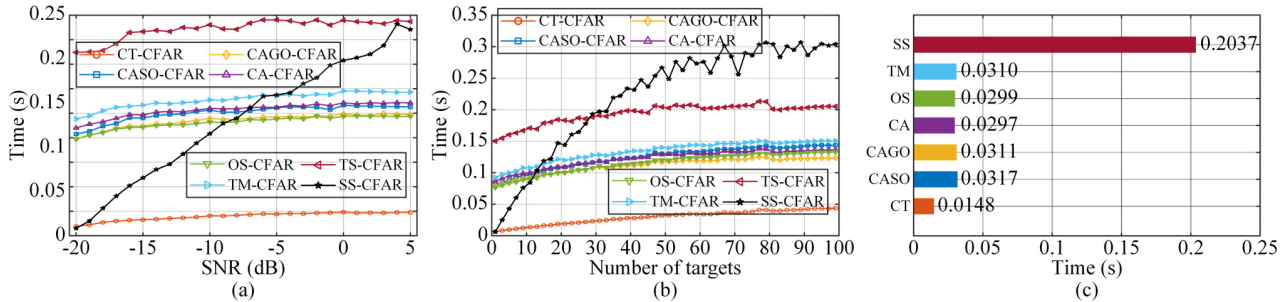

**Fig. 13.** Runtime analysis of different CFAR algorithms. (a) Runtime vs. SNR at $P_{FA}=10^{-3}$. (b) Runtime vs. number of targets at $P_{FA}=10^{-3}$ and SNR=0dB. (c) Runtime in measurement experiments.

Fig. 12 shows the radar point cloud results obtained using CT-CFAR and several comparative CFAR algorithms as target detectors, following angle-of-arrival estimation. The magnified details of the human body region are highlighted by the boxes in the figure.

The experimental results indicate that CT-CFAR and SS-CFAR achieve the best overall detection performance, exhibiting strong suppression of target masking effects and sidelobe-induced false alarms, thereby more accurately reconstructing human body contours. In terms of local details, CT-CFAR provides spatial resolution comparable to CASO-CFAR and OS-CFAR, and, compared to SS-CFAR, is better able to preserve the full features of target postures.

*D. Runtime Performance Evaluation*

To evaluate the computational efficiency of various CFAR algorithms, we analyze their program runtime using both simulation and actual measurement experiments. Fig. 13 shows the comparison of the execution times for different CFAR algorithms.

Specifically, Fig. 13 (a) shows the variation of runtime with SNR in the simulation experiments. The runtimes of SS-CFAR and TS-CFAR increase markedly as SNR rises, with SS-CFAR showing the most significant growth, while the runtimes of the other detection algorithms remain largely unaffected by changes in SNR. Overall, CT-CFAR consistently achieves the highest computational efficiency across all SNR levels. Fig. 13 (b) shows the runtime as a function of the number of targets in the simulated experiments. As the number of targets increases, SS-CFAR exhibits a steep rise in runtime, whereas the other algorithms show only minor increases, keeping computational costs within an acceptable range. Fig. 13 (c) shows the runtime results obtained from measured data. Consistent with the simulation findings, CT-CFAR maintains superior execution efficiency in real-world experiments as well.

V. CONCLUSION

In this paper, we propose CT-CFAR to address the non-homogeneity of the reference window caused by sidelobe contamination. Unlike mainstream mean-based, order-statistic, and truncated CFAR variants, CT-CFAR does not rely on prior knowledge of anomalous samples; instead, it adaptively separates target and noise spectra to restore the homogeneity assumption of the reference window. The proposed method achieves this while preserving detection capability and estimation accuracy, and it additionally offers superior computational efficiency.

The proposed CT-CFAR algorithm achieves precise separation between target and noise spectra through TS, restoring the optimal detection conditions of mean-based CFAR algorithms while incorporating learnable historical sidelobe information to enhance robustness. Multi-antenna observations are used to reconstruct target echoes via the Candan algorithm and LS estimation, and the integration of the CLEAN concept effectively suppresses sidelobe interference. Monte Carlo simulations and real-world measurements validate that CT-CFAR accurately separates target and noise components and significantly reduces false alarms. Compared with mainstream mean-based, order-statistic, and advanced CFAR algorithms such as TS-CFAR and SS-CFAR, CT-CFAR demonstrates superior detection capability, accuracy, and computational efficiency. Nevertheless, CT-CFAR exhibits limited detection performance when multiple targets are closely spaced, with inter-target distances on the order of 1-2 range resolution cells. Future work will focus on modeling and joint detection of closely spaced targets to further enhance the algorithm's performance in dense target scenarios.